\documentclass[a4paper]{article}

\usepackage{INTERSPEECH2019}

\usepackage{multirow}
\usepackage{amsmath,amssymb}
\usepackage[affil-it]{authblk}
\usepackage[usenames, dvipsnames]{color}
\usepackage{enumitem}
\usepackage{booktabs}
\usepackage{tabu}
\usepackage[usenames,dvipsnames]{xcolor}
\usepackage{footnote}
\usepackage{ctable} % for \specialrule command
\makeatletter
\def\thickhline{%
  \noalign{\ifnum0=`}\fi\hrule \@height \thickarrayrulewidth \futurelet
   \reserved@a\@xthickhline}
\def\@xthickhline{\ifx\reserved@a\thickhline
               \vskip\doublerulesep
               \vskip-\thickarrayrulewidth
             \fi
      \ifnum0=`{\fi}}

\renewcommand\section{\@startsection {section}{1}{\z@}%
                                    {-1.5ex \@plus 0.0ex \@minus -.2ex}%
                                    {1.5ex \@plus.12ex}%
                                    {\normalfont\Large\bfseries\color{mybrown}}}
\renewcommand\subsection{\@startsection{subsection}{2}{\z@}%
                                      {-1.5ex \@plus 0.0ex \@minus -.2ex}%
                                      {0.75ex \@plus 0.12ex}%
                                      {\normalfont\large\bfseries\color{mybrownlighter}}}
\renewcommand\subsubsection{\@startsection{subsubsection}{3}{\z@}%
                                          {-0.25ex\@plus -0.1ex \@minus -.2ex}%
                                          {0.8ex \@plus .12ex}%
                                          {\normalfont\normalsize\bfseries\color{mybrown}}}
\renewcommand\paragraph{\@startsection{paragrcaph}{5}{\z@}%
                                      {1ex \@plus1.2ex \@minus.75ex}% 
                                      {0em}%
                                      {\normalfont\large\bfseries\slshape\color{mybrown}}}
\renewcommand\subparagraph{\@startsection{subparagraph}{6}{\parindent}%
                                         {0.25ex \@plus0.2ex \@minus .2ex}%
                                         {-0.2em}%
                                         {\normalfont\normalsize\color{mybrown}}}

\renewcommand{\vec}[1]{\mathbf{#1}}
 \makeatother

\newlength{\thickarrayrulewidth}
\title{Speaker Diarization with Lexical Information}
\name{Tae Jin Park$^1$, Kyu J. Han$^2$, Jing Huang$^2$, Xiaodong He$^2$, Bowen Zhou$^2$, Panayiotis Georgiou$^1$ and Shrikanth Narayanan$^1$}
\address{
  $^1$University of Southern California\\
  $^2$JD AI Research}
\email{taejinpa@usc.edu}

\begin{document}

\maketitle

\begin{abstract}
This work presents a novel approach for speaker diarization to leverage lexical information provided by automatic speech recognition. We propose a speaker diarization system that can incorporate word-level speaker turn probabilities with speaker embeddings into a speaker clustering process to improve the overall diarization accuracy. To integrate lexical and acoustic information in a comprehensive way during clustering, we introduce an adjacency matrix integration for spectral clustering. Since words and word boundary information for word-level speaker turn probability estimation are provided by a speech recognition system, our proposed method works without any human intervention for manual transcriptions. We show that the proposed method improves diarization performance on various evaluation datasets compared to the baseline diarization system using acoustic information only in speaker embeddings.  
\end{abstract}
\section{Introduction}
\label{sec:intro}
Speaker diarization is a process of partitioning a given multi-speaker audio signal in terms of ``who spoke when'', generally consisting of two sub-processes: \textit{speaker segmentation} of cutting the given audio into homogeneous speech segments in terms of speaker characteristics and \textit{speaker clustering} of grouping all the segments from one speaker into the same cluster and assigning them with the same speaker label. Speaker diarization plays a critical role in speech applications like automatic speech recognition (ASR) or behavioral analytics \cite{liu2005online,hain2012transcribing,bs1,bs2}. 

Speaker diarization has long been considered a pre-processing step in the context of ASR. This is mostly because, considering research setups where oracle results for speech activity detection or segmentation are given, grouping speech portions from the same speakers in a multi-speaker audio signal can benefit ASR systems. It can enable speaker-specific feature transformation, e.g., fMLLR \cite{gales97} or total variability factor analysis for i-vectors \cite{dehak2011front}. However, such oracle results would not be available for speaker diarization in practice. Also, performing speaker diarization before ASR on production systems in the wild without proper post-processing would degrade recognition accuracy significantly since it is likely to determine speaker change points in the middle of words, not between words, and result in word cuts or deletions. These practical issues were also pointed out in \cite{wordcut}. In addition, we recently showed in \cite{park2018multimodal} that lexical cues in words or utterances can help diarization accuracy improve when combined with acoustic features. All of these suggest that it would make more sense and practical for speaker diarization to be considered as a \textit{post-processing} step and take advantage of utilizing ASR outputs within the ASR pipeline. With this regard, in this paper, we assume that there are available ASR outputs in a text form for speaker diarization and propose a system to incorporate such lexical information into the diarization process.

There have been a handful of works to employ ASR outputs to enhance speaker diarization systems, but mostly limited to speaker segmentation. ASR outputs are used in \cite{cerva2013speaker} for determining potential speaker change points. In \cite{india2017lstm}, the lexical information provided by an ASR system is utilized to train a character-level language model and improve speaker segmentation performance. In our previous work \cite{park2018multimodal}, we exploited lexical information, from either reference transcripts or ASR outputs, along with acoustic information to enhance speaker segmentation in estimating speaker turns and showed the overall improvement in speaker diarization accuracy. 

In this paper we extend the exploitation of lexical information provided by an ASR system to a \textit{speaker clustering} process in speaker diarization. The challenge of employing lexical information to speaker clustering is multifaceted and requires practical design choices. In our proposal, we use \textit{word-level speaker turn probabilities} as lexical representation and combine them with acoustic vectors of \textit{speaker embedding} when performing \textit{spectral clustering} \cite{von2007tutorial}. In order to integrate lexical and acoustic representations in the spectral clustering framework, we create \textit{adjacency matrices representing lexical and acoustic affinities between speech segments respectively} and combine them later with a per-element max operation. It is shown that the proposed speaker diarization system improves a baseline performance on two evaluation datasets. 

\begin{figure}[t]
  \centerline{\includegraphics[width=8.0cm]{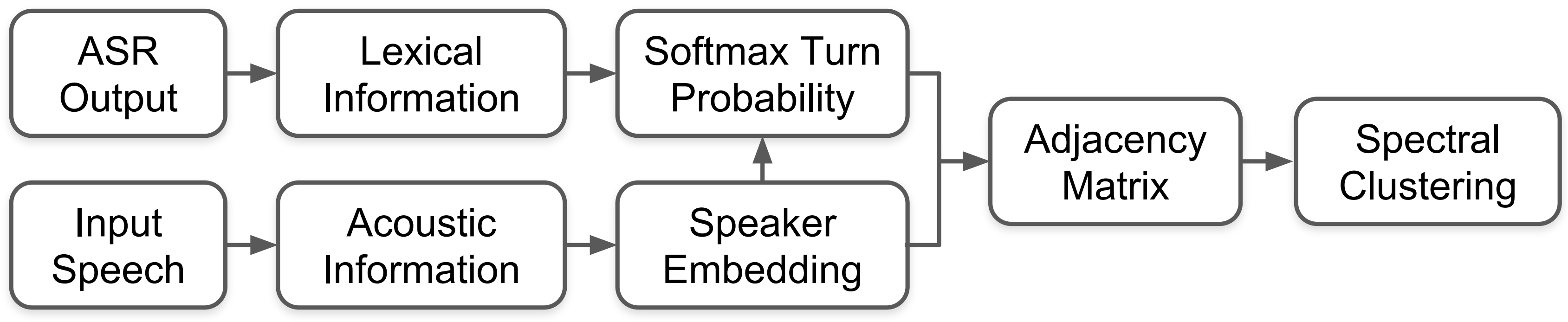}}
  \vspace{-1ex}
  \caption{The data flow of the proposed system.}
\label{fig:data_flow}
\vspace{-5ex}
\end{figure}

The rest of the paper is organized as follows. In Section \ref{sec:dataflow_of_the_diarization_system}, we explain the data flow of our proposed speaker diarization system. In Sections \ref{sec:acoustic_info} and \ref{sec:lexical_info}, we detail how we process acoustic and lexical information, respectively. In Section \ref{sec:adjacency_matrix_integration}, we describe the integration of the two sets of information in the framework of spectral clustering. Experimental results are discussed in Section \ref{sec:Experimental Results} and we conclude the work with some remarks in Section \ref{sec:conclusions}.

\section{Proposed speaker diarization system}
\label{sec:dataflow_of_the_diarization_system}
% \boldred{How our system works}
The overall data flow of our proposed speaker diarization system is depicted in Fig \ref{fig:data_flow}. In the proposed system, there are two streams of information: lexical and acoustic. On the lexical information side, we receive the automated transcripts with the corresponding time stamps for word boundaries from an available ASR system. This text information is passed to the speaker turn probability estimator to compute word-level speaker turn probabilities. On the acoustic information side, we perform a common diarization task. MFCCs are extracted from the input speech signal after speech activity detection (SAD). Following SAD, we uniformly segment the SAD outputs. These uniform segments are relayed to the speaker embedding extractor that provides speaker embedding vectors. We use the publicly available Kaldi ASpIRE SAD Model\footnote{http://kaldi-asr.org/models/m4} \cite{povey2011kaldi} for SAD in our proposed diarization pipeline. 
% We use 20-dimensional MFCCs with a 25ms window with a 10ms shift. 

%Our diarization system is based on fixed-length segments with a 1-second window and a 0.3-second shift.
After processing the two streams of information, we create two adjacency matrices which hold lexical as well as acoustic affinities between speech segments, respectively, and combine them with a per-element max operation to generate the combined affinity matrix that contains lexical and acoustic information in a comprehensive space. With the integrated adjacency matrix, we finally obtain speaker labels using a spectral clustering algorithm. Each module in Fig. \ref{fig:data_flow} is explained in more details in the following sections.

% \begin{figure}[t]
%   \centerline{\includegraphics[width=9.0cm]{ebd_extr.png}}
%   \caption{An illustration of the proposed speaker embedding extractor.}
%   \vspace{-4ex}
% \label{fig:ebd_extr}
% \end{figure}
\section{Acoustic information stream:\\ Speaker embedding extractor}  
\label{sec:acoustic_info}
We employ the x-vector model\footnote{http://kaldi-asr.org/models/m6} \cite{snyder2018xvector} as our speaker embedding generator that has been showing the  state-of-the-art performances for speaker verification and diarization tasks. To perform the general diarization task with acoustic information in the proposed system pipeline, we use 0.5 second window, 0.25 second shift and  0.5 second minimum window size for 23-dimensional MFCCs. The performance improvement of speaker embedding is out of the scope of this paper.

\section{Lexical information stream:\\Speaker turn probability estimator}
\label{sec:lexical_info}
While acoustic speaker characteristics can be used for speaker turn detection tasks \cite{bredin2017tristounet}, our proposal of word-level speaker turn probability estimation comes behind the reasoning that lexical data can also provide an ample amount of information for similar tasks. It is likely for words in a given context (i.e., utterance) to have different probabilities on whether speaker turns change at the time of being spoken. For example, the words in the phrase ``how are you'' are very likely to be spoken by a single speaker rather than by multiple speakers. This means that each word in this phrase ``how are you'' would likely have lower speaker turn probabilities than the word right after the phrase would have. In addition to lexical information, we also fuse a speaker embedding vector per each word to increase the accuracy of the turn probability estimation. 

\begin{figure}[t]
%   \centering
  \centerline{\includegraphics[width=6.0cm]{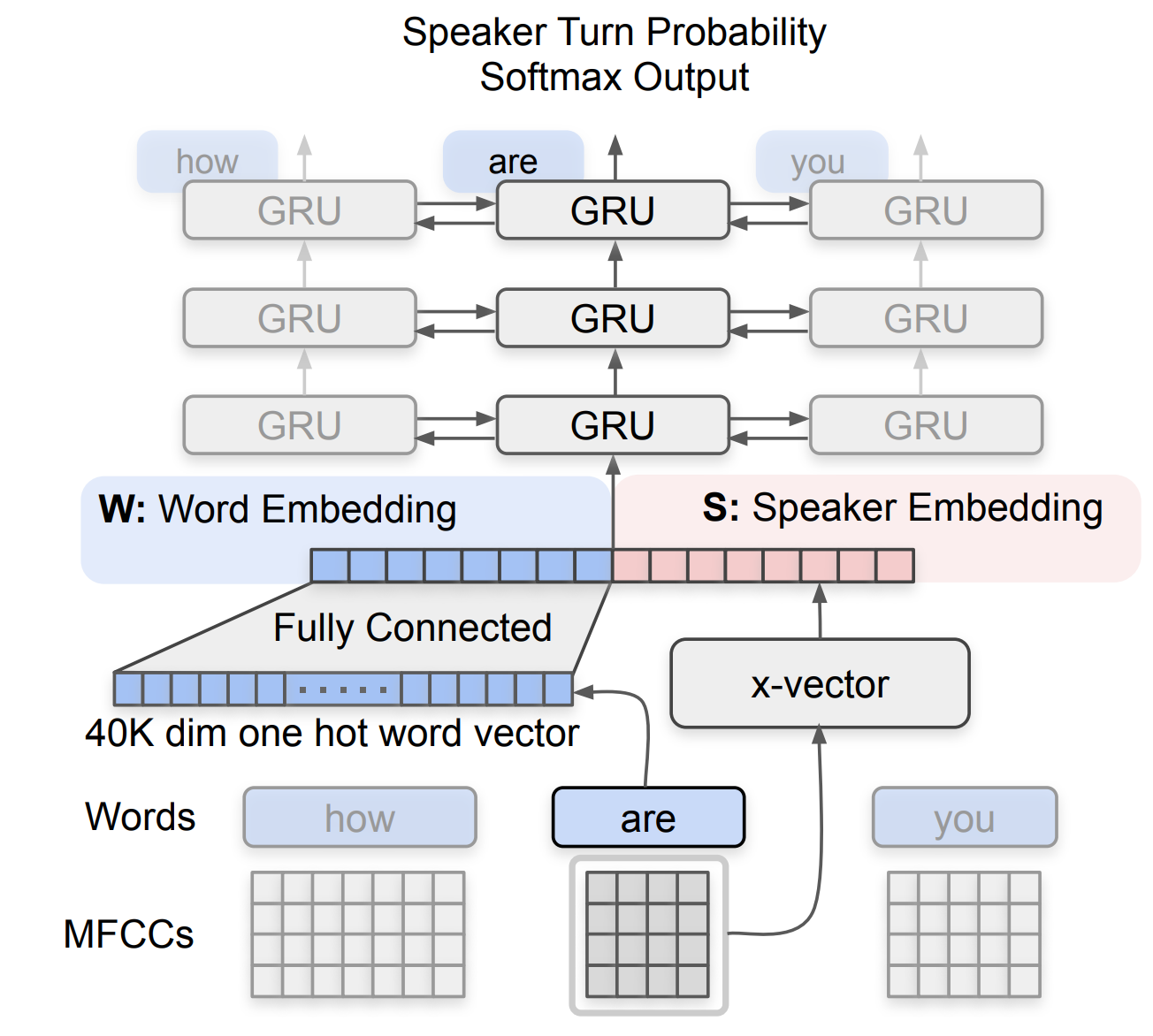}}
  \vspace{-2.0ex}
  \caption{An illustration of the proposed speaker turn probability estimator.}
\label{fig:turn_seg}
\vspace{-4ex}
\end{figure}

To estimate speaker turn probability, we train bi-directional three-layer gated recurrent units (GRUs) \cite{chung2014empirical} with 2,048 hidden units on the Fisher \cite{cieri2004fisher} and Switchboard \cite{godfrey1997switchboard} corpora using the force-aligned texts. The actual inputs to the proposed speaker turn probability estimator would be the decoder outputs of the ASR. The words and the corresponding word boundaries are used to generate word embedding and speaker embedding vectors respectively, as follows: 
\begin{itemize}[leftmargin=0.1in,topsep=2pt,itemsep=3pt,partopsep=0pt, parsep=2pt]
    \item \textbf{Speaker embedding vector (S)}: With the given start and end time stamps of a word from ASR, we retrieve the speaker embedding vector using the speaker embedding extractor described in Section  \ref{sec:acoustic_info}. The x-vector speaker embedding is 128-dimensional.
    \item \textbf{Word embedding vector (W)}: We map the same word input to a 40K-dimensional one-hot vector, which is fully connected to the word embedding layer shown in Fig. \ref{fig:turn_seg}. The dimension of the embedding layer is set to 256. 
\end{itemize}
These two vectors are appended to make a 384-dimensional vector for every word and fed to the GRU layer. The softmax layer has one node and, during inference, outputs speaker turn probability. The parameters of the speaker turn probability estimator are trained with the cross entropy loss.
The ASR system used in this paper for decoding
is the Kaldi ASpIRE recipe\footnote{http://kaldi-asr.org/models/m1} \cite{povey2011kaldi} that is publicly available.
% The ASR system used in this paper for the decoding processes is the CNN-bLSTM model with the PronLex phoneset described in \cite{han2018is}, providing the best single system result on the Switchboard testset of 5.6\% WER (Word Error Rate).

\begin{figure*}[ht]
\centering
  \centerline{\includegraphics[width=15.0cm]{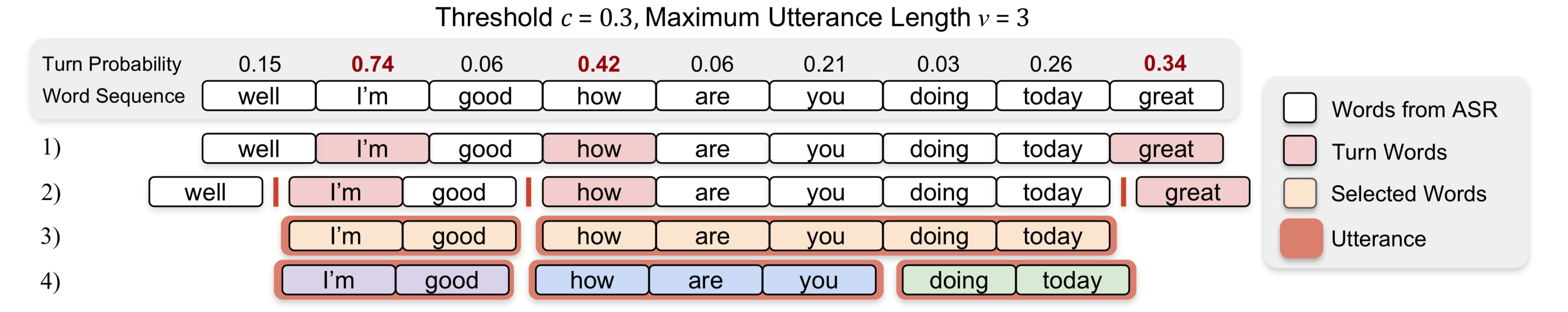}}
  \vspace{-2ex}
  \caption{An example of the word sequence processing for the adjacency matrix calculation using the speaker turn probabilities.}
\label{fig:word_cut}
\vspace{-3ex}
\end{figure*}

\begin{figure}[t]
  \centerline{\includegraphics[width=8.0cm]{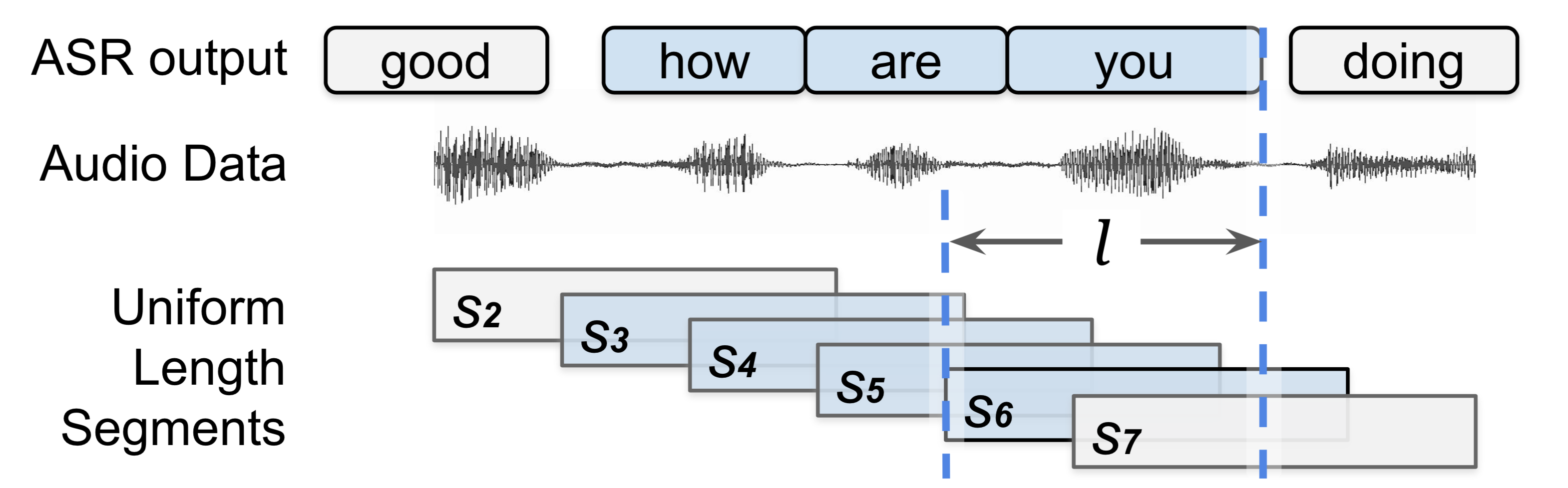}}
  \vspace{-1ex}
  \caption{An example of the speech segment selection process using the utterance boundary information.}
\label{fig:wb_sel}
\vspace{-4.0ex}
\end{figure}

\section{Adjacency matrix integration}
\label{sec:adjacency_matrix_integration}
\subsection{Adjacency matrix calculation}
\label{sec:Adjacency matrix calculation}
% \boldred{the problem of heterogeneous information} 
The biggest challenge of integrating speaker turn probabilities (from lexical information) and speaker embedding vectors (from acoustic information) in the spectral clustering framework is the heterogeneity of the information sources for these representations. To tackle this challenge, we first create two independent adjacency matrices that contain lexical and acoustic affinities between speech segments, respectively, and then combine them with a per-element max operation to handle the information from the two different sources in the common space used for spectral clustering. For each adjacency matrix, we employ undirected graphs to represent the corresponding affinities between the segments. 

% \footnote{In the proposed diarization system of this paper, we apply uniform segmentation of 0.5s-long windowing with a 0.25s shift to a given audio signal. Thus each speech segment here means a 1s-long audio chunk.}

\begin{itemize}[leftmargin=0.1in,topsep=2pt,itemsep=3pt,partopsep=3pt, parsep=3pt]
\item{\textbf{Adjacency matrix using speaker embeddings}}
\begin{enumerate}[leftmargin=0.1in,label=\arabic*),topsep=0pt,itemsep=3pt,partopsep=0pt, parsep=0pt] 
\item Initially compute the cosine similarity $p_{i,j}$ between speaker embedding vectors for segments $s_i$ and $s_j$ to form the adjacency matrix $\mathbf{P}$, where $1 \leq i, j \leq M$ and $M$ is the total number of segments in a given audio signal.

\item For every $i$-th row of $\mathbf{P}$, update $p_{i,j}$ as follows:
\begin{equation}
p_{i,j} =
  \begin{cases}
    1 & \text{if  $p_{i,j} \leq W(r) $}\\
    0 & \text{otherwise}
  \end{cases}
\end{equation}
where $W(r)$ is the cosine similarity value that is at $r$-percentile in each row and $r$ is optimized on the dev set. This operation converts $\mathbf{P}$ to a discrete-valued affinity matrix through $N$ nearest neighbor connections. 

% \footnote{In our experiments, the number of segments ($M$) in audio inputs for the system is 2K on average. Thus, $N$ = 25 means that for each segment we choose the top 1\% among the closest neighboring segments in terms of $L_2$-distance of speaker embeddings.}. 
\item Note that $\mathbf{P}$ is asymmetric and can be seen an adjacency matrix for a directed graph where each node represents a speech segment in our case. As spectral clustering finds the minimum cuts on an \textit{undirected} graph in theory \cite{von2007tutorial}, we choose an undirected version of $\mathbf{P}$, $\mathbf{P_{ud}}$, as the adjacency matrix for speaker embeddings by averaging $\mathbf{P}$ and $\mathbf{P^T}$ as below: 
\begin{equation}
\mathbf{P_{ud}} =\frac{1}{2}(\mathbf{P} + \mathbf{P^T})
\end{equation}
The pictorial representation of $\mathbf{P_{ud}}$ is given in the left side of Fig. 5.
\end{enumerate}

\item{\textbf{Adjacency matrix using speaker turn probabilities}}

%Fig. \ref{fig:word_cut} depicts how lexical information is processed in the softmax output of the speaker turn probabilty estimator in Section \ref{sec:lexical_info} and a word sequence we obtain from an ASR decoder. 
The following steps 1) to 4) match to the numbered illustrations in Fig. \ref{fig:word_cut}, where $c$ = 0.3 and $\nu$ = 3 are given as example parameters. 
% \begin{enumerate}[label=\arabic*)]
\begin{enumerate}[leftmargin=0.1in,label=\arabic*),topsep=2pt,itemsep=1pt,partopsep=1pt, parsep=0pt] 
\item For a given threshold \textit{c}, pick all the \textit{turn words} that have speaker turn probabilities greater than \textit{c} in the word sequence provided by ASR. The colored boxes in Fig. \ref{fig:word_cut}-1) indicate the turn words. The threshold \textit{c} is determined by the eigengap heuristic that we will discuss in Section 5.2.

\item Break the word sequence at every point where the turn word starts as in Fig. \ref{fig:word_cut}-2). The given word sequence is broken into multiple utterances as a result.
\item Pick all the utterances that have more than one word because a duration spanning one word may be too short to carry any speaker-specific information. In Fig. \ref{fig:word_cut}-3), the words ``well" and ``great" are thus not considered for further processing. Additionally, we always arrange seven back channel words (``yes", ``oh", ``okay", ``yeah", ``uh-huh", ``mhm", ``[laughter]") as independent utterances regardless of their turn probabilities.

\item To mitigate the effect of any miss detection by the speaker turn probability estimator, we perform over-segmentation on the utterances by limiting the max utterance length to $\nu$. In Fig. \ref{fig:word_cut}-4), the resulting utterances are depicted with different colors. Maximum utterance length $\nu$ is optimized on the dev set in the range of 2 to 9. 

\item Find all the speech segments that fall into the boundary of each utterance. Fig. \ref{fig:wb_sel} explains how speech segments within the boundary of the example utterance ``how are you'' are selected. If a segment partly falls into the utterance boundary and its overlap ($l$ in Fig. \ref{fig:wb_sel}) is greater than 50\% of the segment length, the segment is considered to fall into the utterance boundary. 
\item Let $s_{m}$ be the first segment and $s_{n}$ be the last segment falling into the utterance boundary (e.g., segments $s_3$ and $s_6$, respectively, in Fig. \ref{fig:wb_sel}). For the elements $q_{i,j}$ in an adjacency matrix \textbf{$\mathbf{Q_{c}}$} (with the threshold $c$) being initialized with zeros, we do the following operation for all the utterances:
\begin{equation}
\begin{split}
% &q_{s_{1}+n,s_{1}+n} = 1 \\
% \text{where} & \;\; 0 \leq n \leq (s_{2}-s_{1})
q_{i,j} =
  \begin{cases}
    1 & \text{if}\;\; m \leq i,j \leq n \\
    q_{i,j} & \text{otherwise}
  \end{cases}
  \end{split}
\end{equation}  
The right side of Fig. \ref{fig:adj_mat} shows an example of \textbf{$\mathbf{Q_{c}}$} by the utterance ``how are you'' in Fig. \ref{fig:wb_sel}.
\end{enumerate}

\begin{figure}[t]
  \centerline{\includegraphics[width=8.0cm]{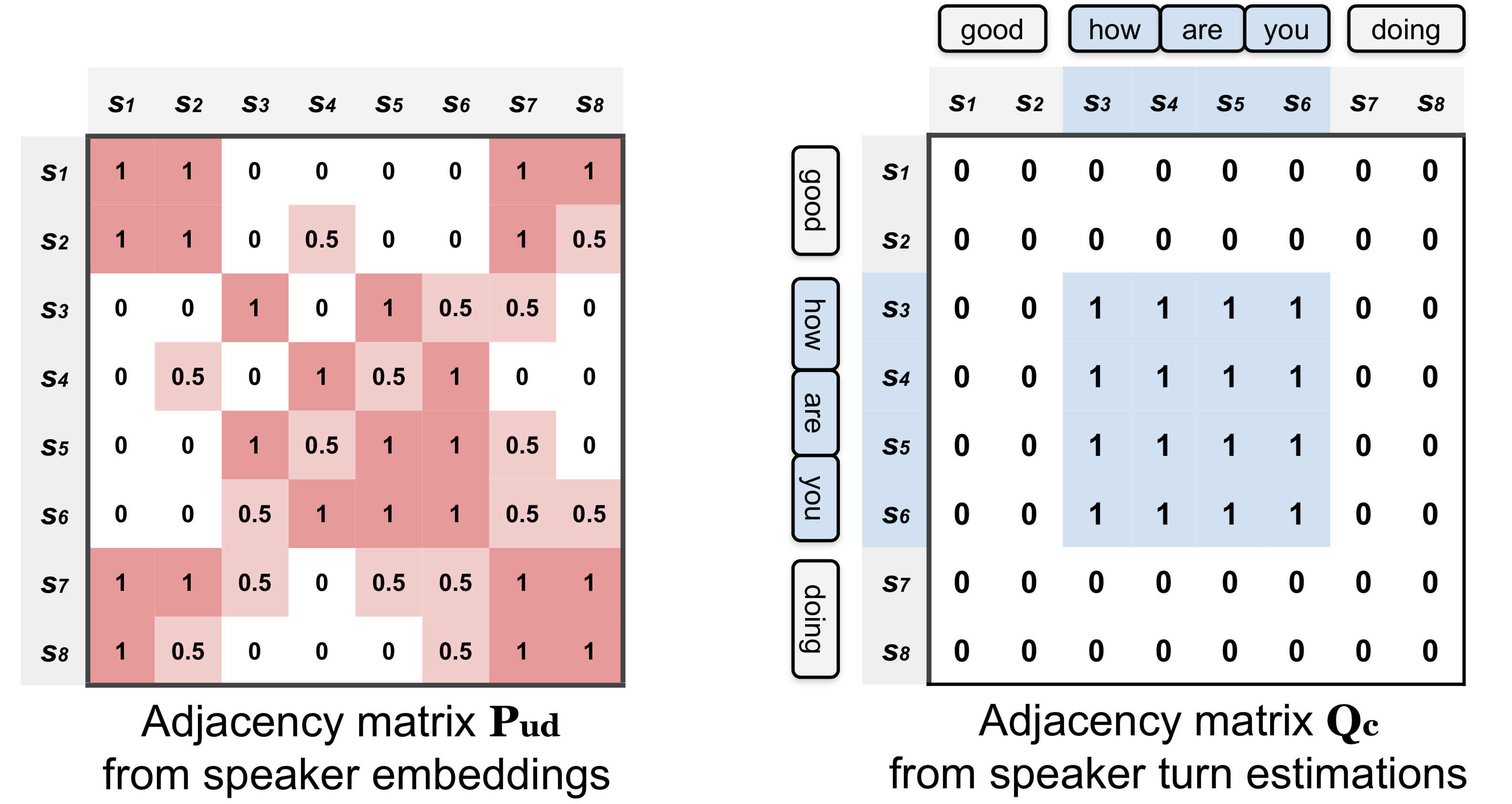}}
\vspace{-1ex}
  \caption{Examples of the two adjacency matrices.}
\label{fig:adj_mat}
\vspace{-4ex}
\end{figure}

\item{\textbf{Combining adjacency matrices}}
\end{itemize}
We combine the two adjacency matrices:
\begin{equation}
\mathbf{A_{c}} =\max{(\mathbf{P_{ud}},\mathbf{Q_{c}})} = \max{(\mathbf{\frac{1}{2}(\mathbf{P} + \mathbf{P}^T),Q_{c}})}
\end{equation}
where $\max$ is a per-element max operation.

\subsection{Eigengap analysis}

In spectral clustering, the Laplacian matrix is employed to get the spectrum of an adjacency matrix. In this work, we employ the unnormalized graph Laplacian matrix $\mathbf{L_{c}}$ \cite{von2007tutorial} as below:
\begin{equation}
\begin{split}
    %d_{i} &= \sum_{k=1}^{M} a_{ik} \\
    %\mathbf{D_{c}} &=\text{diag}\{d_{1}, d_{2}, ..., d_{M}\} \\
    \mathbf{L_{c}} &= \mathbf{D_{c}} - \mathbf{A_{c}}
\end{split}
\end{equation}
where $\mathbf{D_{c}} =\text{diag}\{d_{1}, d_{2}, ..., d_{M}\}$, $d_{i} = \sum_{k=1}^{M} a_{ik}$ and $a_{ij}$ is the element in the $i^{\text{th}}$ row and $j^{\text{th}}$ column of the adjacency matrix $\mathbf{A_{c}}$. We calculate eigenvalues from $\mathbf{L_{c}}$ and set up an eigengap vector $\mathbf{e_{c}}$: 
\begin{equation}
\label{eq:eig_vec}
    \mathbf{e_{c}} = [\lambda_{2}-\lambda_{1}, \lambda_{3}-\lambda_{2}, \cdots, \lambda_{M} -\lambda_{M-1} ]
\end{equation}
where $\lambda_{1}$ is the smallest eigenvalue and $\lambda_{M}$ is the largest eigenvalue. The resulting adjacency matrix $\mathbf{A_{c}}$ is passed to the spectral clustering algorithm, for which we use the implementation in \cite{scikit-learn}. 

The number of clusters (in our case, number of speakers) is estimated by finding the $\arg\max{}$value of the eigengap vector $\mathbf{e_{c}}$ as in the following equation:
\begin{equation}\label{eq:n_s}
    \widehat{n_s} = \underset{n}{\arg\max}{(\mathbf{e_c})}
\end{equation}
where $\widehat{n_s}$ refers to the estimated number of speakers. 

\begin{table}[t]
\centering
\caption{DER (\%) on the RT03-CTS dataset.}
\vspace{-2.0ex}
\label{table:rt03}
\begingroup
\renewcommand{\arraystretch}{1.1} % Default value: 1
\begin{tabular}{ c  l | c c | c c }
\thickhline
% \multicolumn{2}{c|}{\multirow{2}{*}{Method}} & \multicolumn{4}{c}{$\widehat{n_{s}}$}  \\ 
\multicolumn{2}{c|}{Number of Speakers} & \multicolumn{4}{c}{Unknown} \\  \hline
\multicolumn{2}{c|}{Dataset Split(Quantity)} & \multicolumn{2}{c|}{\textbf{Dev}(14)} & \multicolumn{2}{c}{\textbf{Eval}(58)} \\ 
\multicolumn{2}{c|}{Error Type} & \textbf{DER} & \textbf{SER} & \textbf{DER} & \textbf{SER}  \\ \thickhline
\multicolumn{2}{c|}{ Quan \textit{et al.} \cite{wang2018speaker} System SAD}  & - & - & 12.3 & 3.76  \\ \hline
\textbf{Baseline}& \textbf{M1}  & 4.00 & 1.03 & 6.97  & 2.90  \\ \tabucline{-2}
\multirow{2}{*}{\textbf{Proposed}}&\textbf{M2} W & 3.97 & 1.00 & 5.19  & 1.93 \\ 
&\textbf{M3} W+S    & 3.79 & 0.82 & 5.11  & 1.85 \\ \thickhline
% \multicolumn{6}{l}{\footnotesize $\widehat{n_{s}}$: Estimated number of speakers.} \\[-0.3em]
\end{tabular}
\endgroup
\vspace{-4.0ex}
\end{table}

\section{Experimental Results and Discussion}
\label{sec:Experimental Results}

\subsection{Datasets}
Evaluation datasets for our proposed system are limited to English speech corpora since the proposed system relies on the English ASR system at the moment. We report the diarization performance on the following corpora: 
 \begin{itemize}[leftmargin=0.10in,topsep=3pt,itemsep=3.0pt,partopsep=0pt, parsep=1pt] 
 \item{\textbf{RT03-CTS} (LDC2007S10):} We use the 14-vs-58 dev and eval split provided by the authors in \cite{wang2018speaker}. All the parameters appear on this paper are optimized on the RT03 dev set. 
 \item{\textbf{CH American English Speech (CHAES)} (LDC97S42) \cite{linguistic1997callhome}:} Note that the CHAES corpus is different from the commonly-used multilingual dataset ``NIST SRE 2000 CALLHOME (LDC2001S97)" which is the superset of the CHAES corpus. Within the CHAES corpus, we use two different subsets. (1) \textbf{CH-Eval}: Evaluation set from CHAES. (2) \textbf{CH-109}: 109 conversations from the CHAES corpus that have 2 speakers only. The CH-109 subset is popularly used such as in \cite{Zajíc2017} when evaluating diarization systems focusing only on the 2 speaker cases.  
\end{itemize}
% % from CALLHOME (LDC97S42) \cite{linguistic1997callhome}. Note that dataset \cite{linguistic1997callhome} is different from multilingual dataset 2000 NIST SRE (LDC20001S97).  To compare our proposed system with the previous study \cite{wang2018speaker}, we use the same sessions (CH-109) for the evaluation. We also use the 20 session eval set given by the corpus as another evaluation dataset (CH-Eval). 

% To optimize the maximum utterance length parameter $\nu$ for the CH-109 eval data, we use the 72 sessions from the 2003 NIST Rich Transcription (LDC2007S10) English CTS (Conversational Telephone Speech) part as a development set, which we refer to as RT03-CTS in this paper. We also performed a cross validation of the RT03-CTS dataset by separating the sessions into 5 different folds where each fold consists of 14-vs-58 sessions (dev vs eval), shown in Table 1. For the parameter tuning of CH-Eval, we use the train and dev set (total 100 sessions) given by the CALLHOME corpus as a development set.
\subsection{Evaluation setups}
We evaluate the proposed system (M3) with the baseline system configuration (M1) on the two evaluation datasets (CH-109 and CH-Eval) as well as the RT03-CTS dataset. To evaluate the systems in terms of diarization error rate (DER) and speaker error rate (SER), we use the \textit{md-eval} software presented in \cite{fiscus2006rich}. The gap between DER and SER orginates from the false alarms and missed detections that are caused by SAD. The systems compared in the tables above are configured in the following manners:
\begin{itemize}[leftmargin=0.15in,topsep=5pt,itemsep=3.0pt,partopsep=0pt, parsep=1pt] 
\item \textbf{M1}: This baseline system configuration only exploits $\mathbf{P_{ud}}$ as $\mathbf{A_{c}}$ for spectral clustering (i.e., $\mathbf{A_{c}}$ = $\mathbf{P_{ud}}$). This is the general speaker diarization system utilizing acoustic information only in speaker embeddings. The results of this system would contrast how much lexical information can contribute to the speaker clustering process to enhance the overall speaker diarization accuracy in M2 and M3.
\item \textbf{M2}: This configuration for the proposed system excludes the speaker embedding part for the speaker turn probability estimator in Fig. 2 to show the contribution of lexical information in the speaker turn probability estimation process.
\item \textbf{M3}: This is the full-blown configuration, as explained throughout this paper.
\end{itemize}

 \subsection{Evaluation results}
 \label{sec:evaluation_results}
 The performance of our proposed system is compared to previously published results \cite{wang2018speaker, Zajíc2017} on the same dataset. However, it should be noted that results in \cite{wang2018speaker} and our proposed system are based on system SAD that is bound to give higher DER than the systems based on oracle SAD. On the other hand, the system in \cite{Zajíc2017} uses oracle SAD which makes DER equal to SER.
 \begin{itemize}[leftmargin=0.15in,topsep=3pt,itemsep=3.0pt,partopsep=0pt, parsep=1pt] 
 \item{\textbf{Table\ref{table:rt03} (RT03-CTS):}} The M3 system improves the performance over M2, but the relative improvements are minimal as compared to the improvements of M2 over M1. This shows that most of the performance gain by the proposed speaker diarization system comes from employing lexical information to the speaker clustering process. 
 \item{\textbf{Table\ref{table:ch} (CH-Eval, CH-109):}} This table compares our proposed speaker diarization system with the recently published results \cite{wang2018speaker, Zajíc2017} on the CHAES datasets. For a fair comparison, we applied the eigengap analysis based speaker number estimation in Eq. (\ref{eq:eig_vec}) only to the CH-Eval dataset while fixing the number of speakers to 2 in the CH-109 dataset (since CH-109 is the chosen set of the CHAES conversations with only 2 speakers). It is shown in the table that our proposed system (M3) outperforms the previously published results in \cite{wang2018speaker, Zajíc2017} on both CH-Eval and CH-109. It is worthwhile to mention that the proposed system did not gain the noticeable improvement in the CH-Eval dataset as compared to the baseline configuration (M1). As for the CH-109 dataset, on the other hand, M3 seems to provide a noticeble jump in SER over M1. Given our observation that in the CH-109 evaluation most of the performance improvement from M1 to M3 was from the worst 10 sessions that the baseline system performed poorly on, we presume that the proposed system improves the clustering results on such challenging data. 
\end{itemize}

\begin{table}[t]
\centering
\caption{DER (\%) on the CHAES dataset.}
\vspace{-2.0ex}
\label{table:ch}
\begin{tabular}{ c l | c c | c c }
 \thickhline 
\multicolumn{2}{c|}{Number of Speakers}  & \multicolumn{2}{c|}{Unknown}  &  \multicolumn{2}{c}{Known} \\ \hline
\multicolumn{2}{c|}{Dataset(Quantity)}  & \multicolumn{2}{c|}{\textbf{CH-Eval}(20)}  & \multicolumn{2}{c}{\textbf{CH-109}(109)} \\ 
\multicolumn{2}{c|}{Error Type} & \textbf{DER} & \textbf{SER} & \textbf{DER} & \textbf{SER}  \\ \thickhline
\multicolumn{2}{c|}{Quan \textit{et al.} \cite{wang2018speaker} System SAD}  & 12.54 & 5.97 & 12.48 & 6.03 \\ 
\multicolumn{2}{c|}{Zajíc \textit{et al.} \cite{Zajíc2017} Oracle SAD}& - & - &  - & 7.84 \\ \hline
\textbf{Baseline}&\textbf{M1}  & 7.00 & 2.94 & 6.42 & 2.13	 \\ \tabucline{-2}
\multirow{2}{*}{\textbf{Proposed}}&\textbf{M2} W  & 7.04 & 2.97 & 5.96 & 1.67  \\ 
&\textbf{M3} W+S  & 6.97 & 2.9 & 6.03 & 1.73 \\ \thickhline
% \multicolumn{6}{l}{\footnotesize $\widehat{n_{s}}$: Estimated number of speakers.} \\[-0.3em]
% \multicolumn{6}{l}{\footnotesize $n_{s}$=2: Number of speakers fixed to 2.}
\end{tabular}
\vspace{-4.0ex}
\end{table}
%\subsection{Discussion} % Word count should be less than 150!!
%\label{sec:discussion}
\subsection{Discussion}

The experimental results show that the baseline system outperforms the previously published results due to the performance of ASpIRE SAD \cite{povey2011kaldi} and x-vector \cite{snyder2018xvector}. However, our proposed system still improves the competitive baseline system by 36\% for RT03-Eval and 19\% for CH-109 in terms of SER. 
\section{Conclusions}
\label{sec:conclusions}
In this paper, we proposed the speaker diarization system to exploit lexical information from ASR to the speaker clustering process to improve the overall DER. The experimental results showed that the proposed system provides  meaningful improvements on both of the CHAES and RT03-CTS datasets outperforming  the baseline system which is already competitive against the previously published state-of-the-art results. This supports our claim that lexical information can improve diarzation results by incorporating turn probability and word boundaries. Further studies should target the optimal approaches of integrating the adjacency matrices by employing improved search techniques which can improve not only the clustering performance but also the processing speed.

\section{Acknowledgements}
This research was supported in part by NSF, NIH, and DOD. 

\bibliographystyle{IEEEtran}

\bibliography{mybib}

\end{document}